\documentclass[11pt]{article}
\usepackage{hyperref}
\newcommand{\HRule}{\rule{\linewidth}{0.5mm}}
\renewcommand*{\thefootnote}{\fnsymbol{footnote}}

\usepackage{graphics,amssymb,float}
\usepackage{graphicx}
\usepackage{caption}
\usepackage{rotating}
\usepackage{dcolumn}
\usepackage{bm}
\usepackage{cite}
\usepackage{amsmath}
\usepackage{indentfirst} 
\usepackage{epstopdf}
\usepackage[usenames,dvipsnames]{xcolor}
\usepackage{setspace}

\def\lsim{\;\raise0.3ex\hbox{$<$\kern-0.75em\raise-1.1ex\hbox{$\sim$}}\;}
\def\gsim{\;\raise0.3ex\hbox{$>$\kern-0.75em\raise-1.1ex\hbox{$\sim$}}\;}
\def\beq{\begin{equation}}   \def\eeq{\end{equation}}
\def\ba{\begin{array}}       \def\ea{\end{array}}
\def\bea{\begin{eqnarray}}   \def\eea{\end{eqnarray}}

\begin{document}
\hypersetup{pageanchor=false}
\thispagestyle{empty}

	\begin{flushright}
	  {\large
            \textbf{\href{https://cds.cern.ch/record/2885897}{LHCHWG-2024-002}}} \\[0.5cm]	
		{\large 	\textrm{February 21, 2024}} \\[1.0cm]
	\end{flushright}

	\begin{center}

	\textsc{\Large 	\href{https://twiki.cern.ch/twiki/bin/view/LHCPhysics/LHCHWG}{LHC Higgs Working Group}\footnote{\href{https://twiki.cern.ch/twiki/bin/view/LHCPhysics/LHCHWG}{\sl https://twiki.cern.ch/twiki/bin/view/LHCPhysics/LHCHWG}}} \\[0.5cm]
	\textsc{\Large 	Public Note} \\[0.5cm]
	
	\HRule \\[0.9cm]
	\textbf{\Large Benchmark Lines and Planes for Higgs-to-Higgs Decays in the NMSSM} \\[0.5cm]
	\HRule \\[0.5cm]

	\textrm{\large
          Conveners of the NMSSM Subgroup of WG3:\\
          %
          Ulrich Ellwanger$^{1,}$\footnote{\href{mailto:ulrich.ellwanger@ijclab.in2p3.fr}{ulrich.ellwanger@ijclab.in2p3.fr}},
          Margarete M\"uhlleitner$^{2,}$\footnote{\href{mailto:margarete.muehlleitner@kit.edu}{margarete.muehlleitner@kit.edu}}, 
          Nikolaos Rompotis$^{3,}$\footnote{\href{mailto:nikolaos.rompotis@cern.ch}{nikolaos.rompotis@cern.ch}},
          Nausheen R. Shah$^{4,}$\footnote{\href{mailto:nausheen.shah@wayne.edu}{nausheen.shah@wayne.edu}}, 
          Daniel Winterbottom$^{5,}$\footnote{\href{mailto:daniel.winterbottom@cern.ch}{daniel.winterbottom@cern.ch}}} \\[0.3cm]
	\textit{$^{1}$ IJCLab,  CNRS/IN2P3, University  Paris-Saclay,\\ 91405  Orsay,  France} \\
	\textit{$^{2}$ Institute for Theoretical Physics, Karlsruhe Institute of Technology,\\ Wolfgang-Gaede-Str. 1, 76131 Karlsruhe, Germany} \\
	\textit{$^{3}$ Oliver Lodge Laboratory, University of Liverpool,\\ Liverpool, L69 7ZE, United Kingdom} \\
        \textit{$^{4}$ Department of Physics and Astronomy, Wayne State University,\\ Detroit, MI 48201, USA} \\
        \textit{$^{5}$ Imperial College London, Physics Department, Blackett Laboratory,\\ Prince Consort Rd, London, SW7 2BW, United Kingdom} 
	\end{center}
        
        \newpage
        \thispagestyle{empty}



\mbox{}\vspace*{3em}
\begin{center}
	\textbf{Abstract}
\end{center}
A number of benchmark scenarios for NMSSM Higgs boson searches via Higgs-to-Higgs decays at the LHC have been proposed by the NMSSM Subgroup of the LHC HWG3. Some of them are already in use by the ATLAS and CMS collaborations for the interpretation of their results from Run 2. In this document we summarize the theory setup, the underlying procedures and reproduce the benchmark scenarios in table form.

\newpage
\hypersetup{pageanchor=true}
\renewcommand*{\thefootnote}{\arabic{footnote}}
\setcounter{footnote}{0}
\section{The Higgs Sector of the NMSSM}

The Next-to-Minimal Supersymmetric Standard Model (NMSSM)~\cite{Maniatis:2009re,Ellwanger:2009dp} solves the $\mu$-problem of the Minimal Supersymmetric Standard Model (MSSM)~\cite{Kim:1983dt} since the supersymmetric mass term of the Higgs doublets is replaced by the vacuum expectation value (vev) of an extra gauge singlet scalar. This vev is naturally of the order of the supersymmetry breaking scale, somewhat larger than the electroweak scale. The neutral Higgs sector of the NMSSM consists of three complex scalars $H_u^0$, $H_d^0$ and $S$ where $H_u^0$ and $H_d^0$ are members of SU(2) doublets and $S$ is a gauge singlet~\cite{Maniatis:2009re,Ellwanger:2009dp}. Their self couplings originate from terms 
\beq\label{eq:w}
W=\lambda {\cal H}_u {\cal H}_d {\cal S} + \frac{\kappa}{3} {\cal S}^3 + \dots
\eeq
in the superpotential $W$ in terms of superfields ${\cal H}$ and ${\cal S}$, and from trilinear soft supersymmetry breaking terms
\beq\label{eq:A}
(\lambda A_\lambda  H_u  H_d  S + \frac{\kappa}{3} A_\kappa S^3) + \text{h.c.}\; .
\eeq
Contributions from $D$-terms are relatively small. For the
phenomenological analysis of this model, the self couplings have to be
expressed in terms of physical states. To this end the weak
eigenstates $H_u^0$, $H_d^0$ and $S$  have to be expanded around their
vacuum expectation values $v_u$, $v_d$ and $v_s$. The mass matrices
have to be diagonalized, and in the CP-conserving case three neutral
CP-even scalars and two neutral CP-odd scalars are obtained (after elimination of the Goldstone bosons). General expressions for these mass matrices including the dominant radiative corrections are given in Ref.~\cite{Ellwanger:2009dp}.

A first approximation to the physical states is obtained in the so-called Higgs basis where singlet-doublet mixing is neglected and the CP-even doublets are rotated by the same angle as the CP-odd sector. Resulting approximate expressions for the physical masses and trilinear couplings among the physical states are given in Refs.~\cite{Carena:2015moc,Ellwanger:2022jtd}; these allow us to understand orders of magnitude and hierarchies among the trilinear couplings.

The three CP-even scalars are denoted by $H_1$, $H_2$ and $H_3$, which are ordered in mass. The role of the observed SM-like Higgs scalar $H_{SM}$ near 125~GeV is played by $H_1$ or $H_2$ depending on whether the mostly singlet-like scalar $H_{S}$ is lighter or heavier than 125~GeV. The heaviest scalar $H_3$ is mostly MSSM-like, i.e. nearly degenerate with a CP-odd scalar (typically $A_2$, with $A_1$ mostly singlet-like) and a charged Higgs scalar. 
The processes considered are of the form $ggF \to H_3 \to H_{SM} + H_S$ and $ggF \to A_2 \to H_{SM} + A_1$ for various decays of $H_{SM}$ and $H_S$,
and focusing on $A_1 \to \gamma\gamma$. The latter loop induced branching fraction can be particularly large once tree level couplings of $A_1$ to quarks and leptons are suppressed.

The cross sections times branching fractions (Xsect$\times$BRs) are
computed, including radiative corrections to masses \cite{Slavich:2020zjv} and trilinear
couplings \cite{Nhung:2013lpa, Muhlleitner:2015dua,Borschensky:2022pfc}, using the code {\sf NMSSMTools} \cite{NMSSMTools}, and for some cases {\sf NMSSMCALC} \cite{Baglio:2013iia}. In addition to the NMSSM-specific parameters ($\lambda$, $\kappa$, $A_\lambda$ and $A_\kappa$) these radiative corrections depend on Yukawa and gauge couplings, and on soft supersymmetry breaking squark, slepton, electroweak gaugino and gluino masses. In  {\sf NMSSMTools}, the $H_3$/$A_2$ Higgs production cross sections via $ggF$ at $\sqrt{s} = 13$~TeV are taken from the TWiki web page \cite{LHCyellowreport} (update in CERN Report4 2016) multiplied by the squared reduced couplings  of $H_3$ or $A_2$ to top quarks.

For a given set of masses of the two involved beyond-the-SM (BSM)
Higgs bosons (one MSSM-like and one singlet-like), the
Xsect$\times$BRs are maximized respecting constraints from previous
searches for Higgs bosons, constraints from the properties of $H_{SM}$
(couplings and mass, where we allow for a 3~GeV theory uncertainty), and constraints from $b$-physics and dark matter detection experiments. Hence regions of the NMSSM parameter space can be identified where searches have the sensitivity to probe the presence of  the two involved BSM Higgs bosons. However, this procedure does not allow us to exclude a given set of masses of these Higgs bosons since the Xsect$\times$BRs depends on additional parameters of the NMSSM.

Constraints from dark matter detection experiments are relevant since the dark matter relic density and detection cross section depend on the same NMSSM parameters as the Higgs sector. All these constraints are implemented in the code {\sf NMSSMTools}~\cite{NMSSMTools} coupled to 
{\sf MicrOmegas}~\cite{Belanger:2013oya}. Clearly, constraints from searches for BSM Higgs bosons and dark matter detection experiments are time dependent. Therefore we indicate for each benchmark plane which version of {\sf NMSSMTools} has been used. The constraints applied in each version and the associated references can be found on the History tab of the {\sf NMSSMTools} web page~\cite{NMSSMToolsWeb}.  
 
Further benchmark points or planes including Higgs pair production in the NMSSM can be found in
Refs.~\cite{Ellwanger:2022jtd,King:2012is,Kang:2013rj,King:2014xwa,Carena:2015moc,Ellwanger:2015uaz,Costa:2015llh,Baum:2017gbj,Ellwanger:2017skc,Baum:2017enm,Basler:2018dac,Baum:2019uzg,Barducci:2019xkq,Biekotter:2021qbc,Abouabid:2021yvw}.

\section{Benchmark Planes}
 
\subsection{\sf $b\bar{b}b\bar{b}$, $b\bar{b}\tau\tau$, $b\bar{b}\gamma\gamma$}

Figure~1 shows the Xsect$\times$BRs for the processes 
$ggF\to H_3\to (H_{SM}\to b\bar{b}/\tau\tau)+(H_S\to \tau\tau/b\bar{b})$, and $ggF\to A_2\to (H_{SM}\to b\bar{b})+(A_1\to \gamma\gamma)$ as function of $M_{H_3}$ or $M_{A_2}$, for $M_{H_S},\ M_{A_1} \sim 100$~GeV. The corresponding numbers are given in Table~1 in the Appendix. The maximally possible Xsect$\times$BRs for $A_{1}+H_{SM} \to\gamma\gamma+b\bar{b} $ can be relatively large since the BR($ A_{1}\to \gamma\gamma $) can be as large as $\sim 90\% $.

\begin{figure}[h!]
\begin{center}
\hspace*{-10mm}
\includegraphics[scale=0.65]{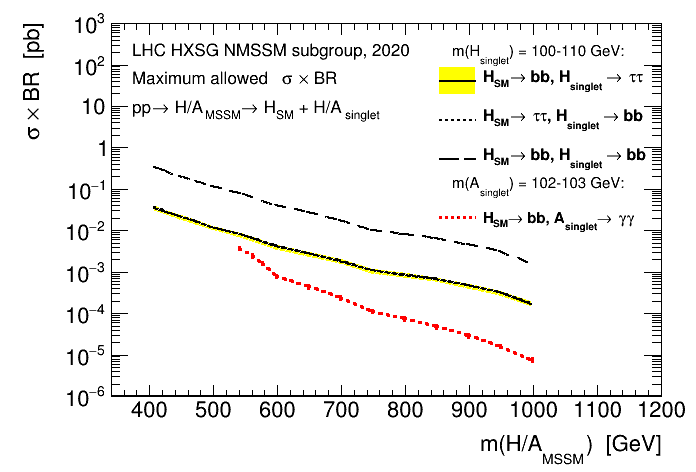}
\end{center}
\vskip -3.mm
\caption{The lines correspond to the maximum cross-section times branching ratio allowed in the (real) NMSSM after experimental constraints from Higgs boson measurements, searches for supersymmetry, B-meson physics and dark matter detection as implemented in the used codes {\sf NMSSMTools\_5.5.0} \cite{NMSSMTools} and {\sf NMSSMCALC} \cite{Baglio:2013iia}. The numbers are given in Table~1 in the Appendix.}
\label{fig:1}
\end{figure}

In Table~2 in the Appendix we show benchmark points for $ggF\to H_{3}\to (H_{SM}\to \gamma \gamma) + (H_{S} \to b\bar{b})$ produced using {\sf NMSSMTools\_5.6.1}. However, constraints from $H/A\to Z+A/H$ by CMS~\cite{CMS:2019ogx} 
or ATLAS~\cite{ATLAS:2020gxx} and from $H/A\to H_{SM}+H/A$ by CMS~\cite{CMS:2021yci} 
are switched off in order to allow comparisons of limits on Xsect$\times$BRs for $(H_{SM}\to \gamma \gamma) + (H_{S} \to b\bar{b})$ obtained by CMS in Ref.~\cite{CMS:2023boe}. These benchmark points have been used in Ref~\cite{CMS:2023boe}.

\subsection{\sf $\tau\tau b\bar{b}$}

In Table~3 in the Appendix we show maximally possible Xsect$\times$BRs for the processes $ggF\to H_3\to (H_{SM}\to \tau\tau)+(H_S\to b\bar{b})$ for $400\text{GeV} \leq M_{H_3} \leq 600 \text{GeV}$, $150\text{GeV} \leq M_{H_S} \leq 300 \text{GeV}$ produced with {\sf NMSSMTools\_5.5.0}. 
These points were used to produce the 2D exclusion region in the CMS publication Ref.~\cite{CMS:2021yci}.

\subsection{\sf $b\bar{b}b\bar{b}/b\bar{b}\tau\tau$}

In Table~4 in the Appendix we show maximally possible Xsect$\times$BRs for the processes $ggF\to H_3\to (H_{SM}\to b\bar{b}/\tau\tau)+(H_S\to b\bar{b})$ for $900\text{GeV} \leq M_{H_3} \leq 3000 \text{GeV}$, $62\text{GeV} \leq M_{H_S} \leq 600 \text{GeV}$ produced with {\sf NMSSMTools\_5.6.1}. 
These points were used to produce the 2D exclusion region in the CMS publication Ref.~\cite{CMS:2022suh}.

\subsection{\sf $\tau\tau\gamma\gamma$}

Table~5 in the Appendix lists maximally possible Xsect$\times$BRs for the processes $ggF\to A_2\to (H_{SM}\to \tau\tau)+(A_1\to \gamma \gamma) $ produced with {\sf NMSSMTools\_5.5.0}. 

\subsection{\sf $\tau\tau WW/ZZ$}

Table~6 in the Appendix lists maximally possible Xsect$\times$BRs for the processes $ggF \to H_{3}\to (H_{SM} \to \tau\tau) + (H_{S} \to WW\ \mbox{or}\ ZZ) $ produced with {\sf NMSSMTools\_6.0.0}. 
These points were used to produce the 2D exclusion region in the ATLAS publication Ref.~\cite{ATLAS:2023tkl}.

\newpage

\appendix
\section*{Appendix}

%
\begin{table}[h!] 
\begin{center}
\caption{
Xsect$\times$BRs for the processes 
$ggF\to H_3\to (H_{SM}\to b\bar{b}/\tau\tau)+(H_S\to \tau\tau/b\bar{b})$, and $ggF\to A_2\to (H_{SM}\to b\bar{b})+(A_1\to \gamma\gamma)$ as function of $M_{H_3}$ or $M_{A_2}$, for $M_{H_S},\ M_{A_1} \sim 100$~GeV. 
The Xsect$\times$BRs for $H_3\to (H_{SM}\to b\bar{b})+(H_S\to b\bar{b})$ and for $(A_{1}+H_{SM}) \to (\gamma\gamma+b\bar{b} )$
in the format $xx/yy$ denote Xsect$\times$BRs obtained by {\sf NMSSMTools\_5.5.0}\cite{NMSSMTools} or {\sf NMSSMCALC}\cite{Baglio:2013iia} respectively (not all mass combinations for $b\bar{b}b\bar{b}$ were obtained by {\sf NMSSMCALC}).
The Xsect$\times$BRs for $H_3\to (H_{SM}\to b\bar{b}/\tau\tau)+(H_S\to \tau\tau/b\bar{b})$ in the format $xx/yy$ denote Xsect$\times$BRs for $H_3\to (H_{SM}\to \tau\tau)+(H_S\to b\bar{b})$ or $H_3\to (H_{SM}\to b\bar{b})+(H_S\to \tau\tau)$ respectively, both obtained by {\sf NMSSMTools\_5.5.0}.
All Xsect$\times$BRs are given in pb.
}
\begin{tabular}{| c | c | c || c | c |}
\hline
$M_{H_3}$ & Xsect$\times$BR for $b\bar{b}b\bar{b}$ & Xsect$\times$BR for $b\bar{b}\tau\tau/\tau\tau b\bar{b}$ & $M_{A_2}$ & Xsect$\times$BR for $b\bar{b}\gamma\gamma$ \\
\hline
408 & $3.44\cdot 10^{-1}/3.86\cdot 10^{-1}$ & $3.67\cdot 10^{-2}/3.54\cdot 10^{-2}$ & 
    540 & $3.79\cdot 10^{-3}$\\
    422 & $2.85\cdot 10^{-1}/3.22\cdot 10^{-1}$ & $ 3.04\cdot 10^{-2}/2.92\cdot 10^{-2}$ &
    560 & $2.52\cdot 10^{-3}$ \\
    447	& $2.09\cdot 10^{-1}/2.35\cdot 10^{-1}$ & $ 2.23\cdot 10^{-2}/2.15\cdot 10^{-2}$ &
    577 & $ 1.63\cdot 10^{-3}$ \\
    472 & $1.57\cdot 10^{-1}$ & $ 1.68\cdot 10^{-2}/1.63\cdot 10^{-2}$ &
    599 & $ 7.95\cdot 10^{-4}$ \\
    497	& $1.15\cdot 10^{-1}$ & $ 1.23\cdot 10^{-2}/1.20\cdot 10^{-2}$ &
    648 & $ 4.46\cdot 10^{-4}$ \\
    547 & $7.30\cdot 10^{-2}/7.14\cdot 10^{-2}$ & $ 7.79\cdot 10^{-3}/7.49\cdot 10^{-3}$ &
    698 & $ 2.41\cdot 10^{-4}$ \\
    597 & $4.02\cdot 10^{-2}$ & $ 4.29\cdot 10^{-3}/4.14\cdot 10^{-3}$ &
    748 & $ 1.14\cdot 10^{-4}$ \\
    647 & $2.72\cdot 10^{-2}/2.63\cdot 10^{-2}$ & $ 2.90\cdot 10^{-3}/2.80\cdot 10^{-3}$ &
    798 & $ 7.48\cdot 10^{-5}$ \\
    697 & $1.79\cdot 10^{-2}$ & $ 1.91\cdot 10^{-3}/1.86\cdot 10^{-3}$ &
    848 & $ 4.85\cdot 10^{-5}$ \\
    747 & $1.03\cdot 10^{-2}$ & $ 1.10\cdot 10^{-3}/1.07\cdot 10^{-3}$ &
    898 & $ 2.93\cdot 10^{-5}$ \\
    797 & $8.23\cdot 10^{-3}/7.19\cdot 10^{-3}$ & $ 8.79\cdot 10^{-4}/8.53\cdot 10^{-4}$ &
    948 & $ 1.63\cdot 10^{-5}$ \\
    847 & $6.47\cdot 10^{-3}/5.29\cdot 10^{-3}$ & $ 6.90\cdot 10^{-4}/6.67\cdot 10^{-4}$ &
    998 & $ 7.56\cdot 10^{-6}$ \\
    897 & $4.59\cdot 10^{-3}/3.56\cdot 10^{-3}$ & $ 4.89\cdot 10^{-4}/4.74\cdot 10^{-4}$ & - & - \\
    947 & $3.11\cdot 10^{-3}/2.50\cdot 10^{-3}$ & $ 3.31\cdot 10^{-4}/3.21\cdot 10^{-4}$ & - & - \\
    997 & $1.64\cdot 10^{-3}/1.83\cdot 10^{-3}$ &  $ 1.74\cdot 10^{-4}/1.69\cdot 10^{-4}$ & - & - \\
\hline
\end{tabular}
\end{center}
\end{table}

\begin{table}[h!] 
\begin{center}
\caption{
Maximally possible Xsect$\times$BRs (in pb) at 13~TeV for
$ggF\to H_{3}\to (H_{SM}\to \gamma \gamma) + (H_{S} \to b\bar{b})$ produced using {\sf NMSSMTools\_5.6.1}.
Constraints from $H/A\to Z+A/H$ by CMS~\cite{CMS:2019ogx} 
or ATLAS~\cite{ATLAS:2020gxx} and from $H/A\to H_{SM}+H/A$ by CMS~\cite{CMS:2021yci} 
are switched off.}
\begin{tabular}{| c | c | c || c | c | c || c | c | c |}
\hline
$M_{H_3}$ & $M_{H_{S}}$ & Xsect$\times$BR  & $M_{H_3}$ & $M_{H_{S}}$ & Xsect$\times$BR & $M_{H_3}$ & $M_{H_{S}}$ & Xsect$\times$BR \\
\hline
 410 & 90 & 0.141$\cdot 10^{-2}$&
 409 & 100 & 0.155$\cdot 10^{-2}$&
 400 & 150 & 0.923$\cdot 10^{-3}$\\
 400 & 200 & 0.109$\cdot 10^{-2}$&
 400 & 250 & 0.141$\cdot 10^{-4}$&
 500 & 90 & 0.522$\cdot 10^{-3}$\\
 500 & 100 & 0.527$\cdot 10^{-3}$&
 500 & 150 & 0.472$\cdot 10^{-3}$&
 500 & 200 & 0.386$\cdot 10^{-3}$\\
 500 & 250 & 0.329$\cdot 10^{-3}$&
 500 & 300 & 0.253$\cdot 10^{-3}$&
 600 & 90  & 0.193$\cdot 10^{-3}$\\
 600 & 100 & 0.194$\cdot 10^{-3}$&
 600 & 150 & 0.190$\cdot 10^{-3}$&
 600 & 200 & 0.182$\cdot 10^{-3}$\\
 600 & 250 & 0.722$\cdot 10^{-4}$&
 600 & 300 & 0.142$\cdot 10^{-3}$&
 600 & 400 & 0.103$\cdot 10^{-4}$\\
 700 & 90 & 0.800$\cdot 10^{-4}$&
 700 & 100 & 0.802$\cdot 10^{-4}$&
 700 & 150 & 0.772$\cdot 10^{-4}$\\
 700 & 200 & 0.795$\cdot 10^{-4}$&
 700 & 250 & 0.758$\cdot 10^{-4}$&
 700 & 300 & 0.502$\cdot 10^{-4}$\\
 700 & 400 & 0.129$\cdot 10^{-4}$&
 700 & 500 & 0.197$\cdot 10^{-5}$&
 800 & 90 & 0.345$\cdot 10^{-4}$\\
 800 & 100 & 0.349$\cdot 10^{-4}$&
 800 & 150 & 0.354$\cdot 10^{-4}$&
 800 & 200 & 0.347$\cdot 10^{-4}$\\
 800 & 250 & 0.236$\cdot 10^{-4}$&
 800 & 300 & 0.212$\cdot 10^{-4}$&
 800 & 400 & 0.635$\cdot 10^{-5}$\\
 800 & 500 & 0.229$\cdot 10^{-5}$&
 800 & 600 & 0.463$\cdot 10^{-6}$&
 900 & 90 & 0.162$\cdot 10^{-4}$\\
 900 & 100 & 0.160$\cdot 10^{-4}$&
 900 & 150 & 0.153$\cdot 10^{-4}$&
 900 & 200 & 0.154$\cdot 10^{-4}$\\
 900 & 250 & 0.130$\cdot 10^{-4}$&
 900 & 300 & 0.126$\cdot 10^{-4}$&
 900 & 400 & 0.216$\cdot 10^{-5}$\\
 900 & 500 & 0.117$\cdot 10^{-5}$&
 900 & 600 & 0.629$\cdot 10^{-6}$&
 900 & 700 & 0.311$\cdot 10^{-7}$\\
 1000 & 90 & 0.749$\cdot 10^{-5}$&
 1000 & 100 & 0.742$\cdot 10^{-5}$&
 1000 & 150 & 0.756$\cdot 10^{-5}$\\
 1000 & 200 & 0.788$\cdot 10^{-5}$&
 1000 & 250 & 0.559$\cdot 10^{-5}$&
 1000 & 300 & 0.507$\cdot 10^{-5}$\\
 1000 & 400 & 0.125$\cdot 10^{-5}$&
 1000 & 500 & 0.615$\cdot 10^{-6}$&
 1000 & 600 & 0.471$\cdot 10^{-6}$\\
 1000 & 700 & 0.220$\cdot 10^{-6}$&
 1000 & 800 & 0.236$\cdot 10^{-7}$&
- & - & -\\
\hline
\end{tabular}
\end{center}
\end{table}

\begin{table}[h!] 
\begin{center}
\caption{
Maximally possible Xsect$\times$BRs (in pb) at 13~TeV for 
$ggF\to H_3\to (H_{SM}\to \tau\tau)+(H_S\to b\bar{b})$ produced with {\sf NMSSMTools\_5.5.0}.
}
\begin{tabular}{| c | c | c || c | c | c |}
\hline
$M_{H_3}$ & $M_{H_{S}}$ & Xsect$\times$BR & $M_{H_3}$ & $M_{H_{S}}$ & Xsect$\times$BR\\
\hline
400  & 	190 	& 0.2870$\cdot 10^{-1}$&
400  & 	250 	& 0.8118$\cdot 10^{-2}$\\
450  & 	190 	& 0.2278$\cdot 10^{-1}$&
450  & 	250 	& 0.1722$\cdot 10^{-1}$\\
500  & 	250 	& 0.1217$\cdot 10^{-1}$&
500  & 	300    	& 0.7850$\cdot 10^{-2}$\\
550  & 	190 	& 0.9352$\cdot 10^{-2}$&
550  & 	250 	& 0.7609$\cdot 10^{-2}$\\
600  & 	150 	& 0.4257$\cdot 10^{-2}$&
600  & 	170 	& 0.5056$\cdot 10^{-2}$\\
\hline
\end{tabular}
\end{center}
\end{table}

\begin{table}[h!] 
\begin{center}
\caption{
Maximally possible Xsect$\times$BRs (in pb) at 13~TeV for 
$ggF\to H_3\to (H_{SM}\to b\bar{b}) + (H_S\to b\bar{b})$ and 
$ggF\to H_3\to (H_{SM}\to \tau\tau) + (H_S\to b\bar{b})$
produced with {\sf NMSSMTools\_5.6.1}.
}
\begin{tabular}{| c | c | c | c || c | c | c | c |}
\hline
$M_{H_3}$ & $M_{H_{S}}$ & $H_{SM}\to b\bar{b}$ & $H_{SM}\to \tau\tau$ & $M_{H_3}$ & $M_{H_{S}}$ & $H_{SM}\to b\bar{b}$ & $H_{SM}\to \tau\tau$  \\
\hline
900  & 	62 	& 0.409$\cdot 10^{-2}$ 	& 0.436$\cdot 10^{-3}$&
900  & 	80 	& 0.408$\cdot 10^{-2}$ 	& 0.434$\cdot 10^{-3}$\\
900  & 	100 	& 0.405$\cdot 10^{-2}$ 	& 0.431$\cdot 10^{-3}$&
900  & 	120 	& 0.412$\cdot 10^{-2}$ 	& 0.438$\cdot 10^{-3}$\\
900  & 	130 	& 0.415$\cdot 10^{-2}$ 	& 0.441$\cdot 10^{-3}$&
900  & 	150 	& 0.412$\cdot 10^{-2}$ 	& 0.447$\cdot 10^{-3}$\\
900  & 	200 	& 0.430$\cdot 10^{-2}$ 	& 0.458$\cdot 10^{-3}$&
900  & 	300 	& 0.340$\cdot 10^{-2}$ 	& 0.361$\cdot 10^{-3}$\\
900  & 	400 	& 0.593$\cdot 10^{-3}$ 	& 0.630$\cdot 10^{-4}$&
900  & 	500 	& 0.330$\cdot 10^{-3}$ 	& 0.351$\cdot 10^{-4}$\\
900  & 	600 	& 0.179$\cdot 10^{-3}$ 	& 0.190$\cdot 10^{-4}$&
1000  & 	62 	& 0.191$\cdot 10^{-2}$ 	& 0.203$\cdot 10^{-3}$\\
1000  & 	80 	& 0.191$\cdot 10^{-2}$ 	& 0.203$\cdot 10^{-3}$&
1000  & 	100 	& 0.189$\cdot 10^{-2}$ 	& 0.201$\cdot 10^{-3}$\\
1000  & 	120 	& 0.193$\cdot 10^{-2}$ 	& 0.205$\cdot 10^{-3}$&
1000  & 	130 	& 0.201$\cdot 10^{-2}$ 	& 0.214$\cdot 10^{-3}$\\
1000  & 	150 	& 0.206$\cdot 10^{-2}$ 	& 0.219$\cdot 10^{-3}$&
1000  & 	200 	& 0.217$\cdot 10^{-2}$ 	& 0.231$\cdot 10^{-3}$\\
1000  & 	300 	& 0.134$\cdot 10^{-2}$ 	& 0.143$\cdot 10^{-3}$&
1000  & 	400 	& 0.341$\cdot 10^{-3}$ 	& 0.362$\cdot 10^{-4}$\\
1000  & 	500 	& 0.171$\cdot 10^{-3}$ 	& 0.182$\cdot 10^{-4}$&
1000  & 	600 	& 0.131$\cdot 10^{-3}$ 	& 0.139$\cdot 10^{-4}$\\
1200  & 	62 	& 0.516$\cdot 10^{-3}$ 	& 0.550$\cdot 10^{-4}$&
1200  & 	80 	& 0.513$\cdot 10^{-3}$ 	& 0.546$\cdot 10^{-4}$\\
1200  & 	100 	& 0.504$\cdot 10^{-3}$ 	& 0.536$\cdot 10^{-4}$&
1200  & 	150 	& 0.492$\cdot 10^{-3}$ 	& 0.527$\cdot 10^{-4}$\\
1200  & 	200 	& 0.547$\cdot 10^{-3}$ 	& 0.582$\cdot 10^{-4}$&
1200  & 	300 	& 0.447$\cdot 10^{-3}$ 	& 0.475$\cdot 10^{-4}$\\
1200  & 	400 	& 0.110$\cdot 10^{-3}$ 	& 0.117$\cdot 10^{-4}$&
1200  & 	500 	& 0.564$\cdot 10^{-4}$ 	& 0.600$\cdot 10^{-5}$\\
1200  & 	600 	& 0.422$\cdot 10^{-4}$ 	& 0.449$\cdot 10^{-5}$&
1400  & 	62 	& 0.161$\cdot 10^{-3}$ 	& 0.171$\cdot 10^{-4}$\\
1400  & 	80 	& 0.168$\cdot 10^{-3}$ 	& 0.179$\cdot 10^{-4}$&
1400  & 	100 	& 0.154$\cdot 10^{-3}$ 	& 0.165$\cdot 10^{-4}$\\
1400  & 	120 	& 0.153$\cdot 10^{-3}$ 	& 0.163$\cdot 10^{-4}$&
1400  & 	130 	& 0.152$\cdot 10^{-3}$ 	& 0.163$\cdot 10^{-4}$\\
1400  & 	150 	& 0.156$\cdot 10^{-3}$ 	& 0.166$\cdot 10^{-4}$&
1400  & 	200 	& 0.140$\cdot 10^{-3}$ 	& 0.149$\cdot 10^{-4}$\\
1400  & 	300 	& 0.129$\cdot 10^{-3}$ 	& 0.137$\cdot 10^{-4}$&
1400  & 	400 	& 0.410$\cdot 10^{-4}$ 	& 0.436$\cdot 10^{-5}$\\
1400  & 	500 	& 0.214$\cdot 10^{-4}$ 	& 0.228$\cdot 10^{-5}$&
1400  & 	600 	& 0.161$\cdot 10^{-4}$ 	& 0.171$\cdot 10^{-5}$\\
1600  & 	62 	& 0.480$\cdot 10^{-4}$ 	& 0.512$\cdot 10^{-5}$&
1600  & 	80 	& 0.473$\cdot 10^{-4}$ 	& 0.504$\cdot 10^{-5}$\\
1600  & 	100 	& 0.439$\cdot 10^{-4}$ 	& 0.467$\cdot 10^{-5}$&
1600  & 	150 	& 0.432$\cdot 10^{-4}$ 	& 0.459$\cdot 10^{-5}$\\
1600  & 	200 	& 0.448$\cdot 10^{-4}$ 	& 0.477$\cdot 10^{-5}$&
1600  & 	300 	& 0.312$\cdot 10^{-4}$ 	& 0.331$\cdot 10^{-5}$\\
1600  & 	400 	& 0.128$\cdot 10^{-4}$ 	& 0.136$\cdot 10^{-5}$&
1600  & 	500 	& 0.639$\cdot 10^{-5}$ 	& 0.679$\cdot 10^{-6}$\\
1600  & 	600 	& 0.469$\cdot 10^{-5}$ 	& 0.499$\cdot 10^{-6}$&
1800  & 	62 	& 0.169$\cdot 10^{-4}$ 	& 0.180$\cdot 10^{-5}$\\
1800  & 	80 	& 0.168$\cdot 10^{-4}$ 	& 0.179$\cdot 10^{-5}$&
1800  & 	100 	& 0.165$\cdot 10^{-4}$ 	& 0.176$\cdot 10^{-5}$\\
1800  & 	150 	& 0.150$\cdot 10^{-4}$ 	& 0.160$\cdot 10^{-5}$&
1800  & 	200 	& 0.154$\cdot 10^{-4}$ 	& 0.164$\cdot 10^{-5}$\\
1800  & 	300 	& 0.820$\cdot 10^{-5}$ 	& 0.872$\cdot 10^{-6}$&
1800  & 	400 	& 0.469$\cdot 10^{-5}$ 	& 0.498$\cdot 10^{-6}$\\
1800  & 	500 	& 0.254$\cdot 10^{-5}$ 	& 0.270$\cdot 10^{-6}$&
1800  & 	600 	& 0.187$\cdot 10^{-5}$ 	& 0.199$\cdot 10^{-6}$\\
2000  & 	62 	& 0.661$\cdot 10^{-5}$ 	& 0.703$\cdot 10^{-6}$&
2000  & 	80 	& 0.659$\cdot 10^{-5}$ 	& 0.700$\cdot 10^{-6}$\\
2000  & 	100 	& 0.662$\cdot 10^{-5}$ 	& 0.704$\cdot 10^{-6}$&
2000  & 	150 	& 0.653$\cdot 10^{-5}$ 	& 0.695$\cdot 10^{-6}$\\
2000  & 	200 	& 0.637$\cdot 10^{-5}$ 	& 0.678$\cdot 10^{-6}$&
2000  & 	300 	& 0.283$\cdot 10^{-5}$ 	& 0.301$\cdot 10^{-6}$\\
2000  & 	400 	& 0.191$\cdot 10^{-5}$ 	& 0.203$\cdot 10^{-6}$&
2000  & 	500 	& 0.123$\cdot 10^{-5}$ 	& 0.130$\cdot 10^{-6}$\\
2000  & 	600 	& 0.106$\cdot 10^{-5}$ 	& 0.113$\cdot 10^{-6}$&
2500  & 	62 	& 0.555$\cdot 10^{-6}$ 	& 0.591$\cdot 10^{-7}$\\
2500  & 	80 	& 0.587$\cdot 10^{-6}$ 	& 0.625$\cdot 10^{-7}$&
2500  & 	100 	& 0.594$\cdot 10^{-6}$ 	& 0.632$\cdot 10^{-7}$\\
2500  & 	150 	& 0.613$\cdot 10^{-6}$ 	& 0.653$\cdot 10^{-7}$&
2500  & 	200 	& 0.664$\cdot 10^{-6}$ 	& 0.707$\cdot 10^{-7}$\\
2500  & 	300 	& 0.524$\cdot 10^{-6}$ 	& 0.558$\cdot 10^{-7}$&
2500  & 	400 	& 0.384$\cdot 10^{-6}$ 	& 0.408$\cdot 10^{-7}$\\
2500  & 	500 	& 0.235$\cdot 10^{-6}$ 	& 0.250$\cdot 10^{-7}$&
2500  & 	600 	& 0.179$\cdot 10^{-6}$ 	& 0.190$\cdot 10^{-7}$\\
3000  & 	150 	& 0.404$\cdot 10^{-7}$ 	& 0.430$\cdot 10^{-8}$&
3000  & 	200 	& 0.238$\cdot 10^{-7}$ 	& 0.254$\cdot 10^{-8}$ \\
\hline
\end{tabular}
\end{center}
\end{table}

\begin{table}[h!] 
\begin{center}
\caption{
Maximally possible Xsect$\times$BRs (in pb) at 13~TeV for
$ggF\to A_2\to (H_{SM}\to \tau\tau)+(A_1\to \gamma \gamma) $ produced with {\sf NMSSMTools\_5.5.0}.
}
\begin{tabular}{| c | c | c ||c | c | c |}
\hline
$M_{A_2}$ & $M_{A_1}$ & Xsect$\times$BR & $M_{A_2}$ & $M_{A_1}$ & Xsect$\times$BR  \\
\hline
410 &	70 	& 0.408$\cdot 10^{-2}$ &
405 &	100 	& 0.885$\cdot 10^{-2}$\\
411 &	170 	& 0.524$\cdot 10^{-2}$&
413 &	200 	& 0.406$\cdot 10^{-2}$\\
500 &	70 	& 0.916$\cdot 10^{-3}$&
500 &	100 	& 0.162$\cdot 10^{-2}$\\
500 &	200 	& 0.126$\cdot 10^{-2}$&
600 &	70 	& 0.214$\cdot 10^{-3}$\\
600 &	100 	& 0.365$\cdot 10^{-3}$&
600 &	200 	& 0.370$\cdot 10^{-3}$\\
700 &	70 	& 0.668$\cdot 10^{-4}$&
700 &	100 	& 0.110$\cdot 10^{-3}$\\
700 &	200 	& 0.118$\cdot 10^{-3}$& - & - & -\\
\hline
\end{tabular}
\end{center}
\end{table}

\begin{table}[h!] 
\begin{center}
\caption{
Maximally possible Xsect$\times$BRs (in pb) at 13~TeV for 
$ggF \to H_{3}\to (H_{SM} \to \tau\tau) + (H_{S} \to WW) $ and
$ggF \to H_{3}\to (H_{SM} \to \tau\tau) + (H_{S} \to ZZ) $  
produced with {\sf NMSSMTools\_6.0.0}.
}
\begin{tabular}{| c | c | c | c || c | c | c | c |}
\hline
$M_{H_3}$ & $M_{H_S}$ & $H_S\to WW$ & $H_S\to ZZ$ & $M_{H_3}$ & $M_{H_S}$ & $H_S\to WW$ & $H_S\to ZZ$  \\
\hline
500 	& 200 	&   0.809$\cdot 10^{-2}$ 	&   0.278$\cdot 10^{-2}$&
500 	& 300 	&   0.420$\cdot 10^{-2}$ 	&   0.186$\cdot 10^{-2}$\\
750 	& 200 	&   0.118$\cdot 10^{-2}$ 	&   0.405$\cdot 10^{-3}$&
750 	& 300 	&   0.587$\cdot 10^{-3}$ 	&   0.261$\cdot 10^{-3}$\\
750 	& 400 	&   0.379$\cdot 10^{-3}$ 	&   0.176$\cdot 10^{-3}$&
750 	& 500 	&   0.208$\cdot 10^{-3}$ 	&   0.994$\cdot 10^{-4}$\\
1000 	& 200 	&   0.114$\cdot 10^{-3}$ 	&   0.395$\cdot 10^{-4}$&
1000 	& 300 	&   0.100$\cdot 10^{-3}$ 	&   0.444$\cdot 10^{-4}$\\
1000 	& 400 	&   0.961$\cdot 10^{-4}$ 	&   0.448$\cdot 10^{-4}$&
1000 	& 500 	&   0.105$\cdot 10^{-3}$ 	&   0.501$\cdot 10^{-4}$\\
1250 	& 200 	&   0.231$\cdot 10^{-4}$ 	&   0.794$\cdot 10^{-5}$&
1250 	& 300 	&   0.232$\cdot 10^{-4}$ 	&   0.103$\cdot 10^{-4}$\\
1250 	& 400 	&   0.256$\cdot 10^{-4}$ 	&   0.119$\cdot 10^{-4}$&
1250 	& 500 	&   0.257$\cdot 10^{-4}$ 	&   0.123$\cdot 10^{-4}$\\
1500 	& 200 	&   0.554$\cdot 10^{-5}$ 	&   0.191$\cdot 10^{-5}$&
1500 	& 300 	&   0.546$\cdot 10^{-5}$ 	&   0.243$\cdot 10^{-5}$\\
1500 	& 400 	&   0.621$\cdot 10^{-5}$ 	&   0.289$\cdot 10^{-5}$&
1500 	& 500 	&   0.652$\cdot 10^{-5}$ 	&   0.311$\cdot 10^{-5}$\\
\hline
\end{tabular}
\end{center}
\end{table}

\end{document}